\documentclass[11pt,a4paper]{amsart}
\usepackage{amsmath}
\usepackage{amssymb}
\usepackage{amsfonts}
\usepackage[toc,page]{appendix}
\usepackage{hyperref}
\usepackage{color}

\def\={\ =\ }

\newcommand{\be}{\begin{equation}}
\newcommand{\ee}{\end{equation}}
\newcommand{\beq}{\begin{equation}}
\newcommand{\eeq}{\end{equation}}
\newcommand{\bea}{\begin{eqnarray}}
\newcommand{\eea}{\end{eqnarray}}

\def\ba{\begin{eqnarray}}
\def\ea{\end{eqnarray}}

\theoremstyle{plain}

\numberwithin{equation}{section}

\input{tcilatex}

\begin{document}
\title[]{Wilson loops and free energies in $3d$ $\mathcal{N}=4$ SYM: exact
results, exponential asymptotics and duality}
\author{}
\author{Miguel Tierz}
\address{Departamento de Matem\'{a}tica, Grupo de F\'{\i}sica Matem\'{a}%
tica, Faculdade de Ci\^{e}ncias, Universidade de Lisboa, Campo Grande, Edif%
\'{\i}cio C6, 1749-016 Lisboa, Portugal.}
\email{tierz@fc.ul.pt}
\maketitle

\begin{abstract}
We show that $U(N)$ $3d$ $\mathcal{N}=4$ supersymmetric gauge theories on $%
S^{3}$ with $N_{f}$ massive fundamental hypermultiplets and with a
Fayet-Iliopoulos (FI) term are solvable in terms of generalized Selberg
integrals. Finite $N$ expressions for the partition function and Wilson loop
in arbitrary representations are given. We obtain explicit analytical
expressions for Wilson loops with symmetric, antisymmetric, rectangular and
hook representations, in terms of Gamma functions of complex argument. The
free energy for orthogonal and symplectic gauge group is also given. The
asymptotic expansion of the free energy is also presented, including a
discussion of the appearance of exponentially small contributions. Duality
checks of the analytical expressions for the partition functions are also
carried out explicitly.
\end{abstract}

\section{Introduction}

The study of supersymmetric gauge theories in compact manifolds has been
considerably pushed forward in recent years, after the development of the
localization method \cite{Pestun:2007rz}, which reduces the original
functional integral describing the quantum field theory into a much simpler
finite-dimensional integral. The task of computing observables in a
supersymmetric gauge theory then typically consists in analyzing a resulting
integral representation, which is normally of the matrix model type. There
is a large number of tools available in the study of matrix models. We will
use here results from random matrix theory and the theory of the Selberg
integral \cite{For}.

We study in particular a three-dimensional $\mathcal{N}=4$ gauge theory on $%
S^{3}$, which consists of a $U(N)$ vector multiplet coupled to $N_{f}$
massive hypermultiplets in the fundamental representation, together with a
Fayet-Iliopoulos (FI) term. The rules of localization in three dimensions 
\cite{Kapustin:2009kz,Hama:2010av}, immediately gives the corresponding
matrix model expression for the partition function%
\begin{equation}
Z_{N_{f}}^{U(N)}=\frac{1}{N!}\int {d^{N}\!\mu }\frac{\ e^{\mathrm{i}\eta
\sum_{i}\mu _{i}}}{\prod_{i}\left( 2\cosh (\frac{1}{2}(\mu _{i}+m))\right)
^{N_{f}}}\ \prod_{i<j}4\sinh ^{2}(\frac{1}{2}(\mu _{i}-\mu _{j})),
\label{mod1}
\end{equation}%
A more general case, also including adjoint hypermultiplets has also been
studied, for example in \cite{Grassi:2014vwa,Assel:2015oxa} and, with
different methods, in \cite{Mezei:2013gqa}. In the case of one adjoint
hypermultiplet, the theory is known to be dual to M-theory on $AdS_{4}\times
S^{7}/N_{f}$, where the quotient by $N_{f}$ leads to an $A_{N_{f}-1}$
singularity. While our results could be extended to this more complicated
model, we focus here on (\ref{mod1}). In what follows, we do not include the
customary $(1/N!)$ factor in (\ref{mod1}) and simply call the partition
function $Z_{N}$. We will comment on that term at the end of the paper, when
discussing dualities.

Other recent exact analytical evaluations of free energies of $3d$
supersymmetric gauge theories can be found in \cite%
{Benvenuti:2011ga,Benini:2017dud}. The models studied there are more
general, but due to remarkable identities satisfied by the double sine
functions that appear in their matrix models, can also be analyzed
analytically. However, we still find it valuable to focus on the simpler (%
\ref{mod1}), relate it to an exact solvable model, and extend the study to
Wilson loops. In turn, this leads to non-trivial aspects of the role of the
FI parameter term and an ensuing discussion of exponentially small
contributions in asymptotic expansions, in relationship with the
classification of these 3d theories as \textit{good}, \textit{ugly} or 
\textit{bad} \cite{Gaiotto:2008ak,Assel:2017jgo,Yaakov:2013fza}.

In addition, notice that, in spite of the apparent simplicity of this model,
even simpler models in 3d, like the Abelian gauge theory studied in \cite%
{Russo:2016ueu}, exhibit rich behavior such as large $N_{f}$ phase
transitions. This non-triviality of the models is in large part due to the
presence of the FI parameter, which always implies an oscillatory Fourier
kernel in the matrix model representation. Concerning (\ref{mod1}), we will
give explicit expressions for the free energy at finite and large $N,$ and
likewise for the average of a Wilson loop in arbitrary representation, which
is given by the matrix integral%
\begin{equation}
\left\langle W_{\lambda }(N)\right\rangle =\frac{1}{Z}\int {d^{N}\!\mu }%
\frac{s_{\lambda }(e^{\mu _{1}},...,e^{\mu _{N}})\ e^{\mathrm{i}\eta
\sum_{i}\mu _{i}}}{\prod_{i}\left( 2\cosh (\frac{1}{2}(\mu _{i}+m))\right)
^{N_{f}}}\prod_{i<j}4\sinh ^{2}(\frac{1}{2}(\mu _{i}-\mu _{j})),
\label{s-mod1}
\end{equation}%
where $Z$ refers to (\ref{mod1}) and $s_{\lambda }$ denotes a Schur
polynomial. Aspects of the analysis of Wilson loops in this theory, in
particular the mirror symmetry between Wilson loops and vortex loops, have
recently appeared in \cite{Assel:2015oxa}. The matrix models have been
considered without FI term in \cite{Grassi:2014vwa,Mezei:2013gqa} and with
one in \cite{Assel:2015oxa}. Localization results for a large class of
correlation functions of local operators in three-dimensional $\mathcal{N}$ $%
=4$ gauge theory on $S^{3}$ are given in \cite%
{Dedushenko:2016jxl,Dedushenko:2017avn}.

The case where all the $N_{f}$ masses are different can also be solved
exactly for the partition function \cite{Benvenuti:2011ga} and has been
studied and used in a number of works \cite{Assel:2012cp,Nishioka:2011dq}.
The solution in that case is based on a Cauchy determinant formula. Thus,
the limiting case, when all masses become equal, is not immediate and it
seems that even the partition function of the model has not been analyzed as
the confluent limit of the expression when all masses are distinct. We show
below how the large $N$ behavior of the free energy has a different leading
behavior when all masses are equal, since the usual leading term $\left(
N^{2}/2\right) \log N$ \cite{Assel:2012cp,Nishioka:2011dq,Lozano:2016wrs}
cancels out. Other, in principle more general versions of the matrix model,
dealing for example with the case where the matter content consists of $%
\mathcal{N}=2$ mass deformation of $\mathcal{N}=3$ hypermultiplets, have
been studied in \cite{Taki:2013opa}, using Cauchy theorem and in the context
of factorization of 3d partition functions \cite{Pasquetti:2011fj}. In this
work, infinite series expressions are obtained and hence asymptotics could
not be obtained with such formulas. In contrast to the previous works, our
explicit analytical expressions will all be in terms of special functions,
G-Barnes functions and Gamma functions, of complex argument.

Notice that in the Abelian case, the theory is exactly solvable, since the
observables then reduce to the evaluation of a very well-known Fourier
transform%
\begin{equation}
Z_{t}\left( \eta \right) =\int_{-\infty }^{\infty }\frac{\ e^{\mathrm{i}\eta
\mu }{d\!\mu }}{\left( \cosh (\mu +m)\right) ^{t}}=\frac{\Gamma \left(
\left( t+i\eta \right) /2\right) \Gamma \left( \left( t-i\eta \right)
/2\right) }{\Gamma \left( t\right) },  \label{Abelian}
\end{equation}%
which follows by immediate identification with Euler's beta integral%
\begin{equation}
\int_{-\infty }^{\infty }\frac{\ ce^{-qcy}{d\!y}}{\left( 1+e^{-cy}\right)
^{p+q}}=\int_{0}^{1}t^{p-1}(1-t)^{q-1}dt:=B(p,q)=\frac{\Gamma \left(
p\right) \Gamma \left( q\right) }{\Gamma \left( p+q\right) },  \label{Euler}
\end{equation}%
with $\func{Re}c>0,$ $\func{Re}p,$ $\func{Re}q\geq 0$ and $\Gamma (p)$ the
Gamma function.

The following exact evaluations, given in terms of a number of Barnes $G$%
-functions, admit the analysis of the observables for different large values
of the parameters, such as $N$, $N_{f}$ and $\eta $, leading to asymptotic
series. The presence of the FI parameter makes the analysis richer, since it
then involves complex arguments $z$ for the Barnes $G$-functions and the
Gamma functions below.

Therefore, the large $z$ behavior of the special functions is required,
involving naturally the whole complex plane, which leads to consideration of
Stokes and anti-Stokes lines and the appearance of exponentially small
contributions when crossing them. We can analyze the observables in these
terms by using the exponential asymptotics of Barnes and Gamma functions 
\cite{Nemes,Nemes2,Pasquetti:2009jg}. Therefore, this theory is suitable to
further explore ideas of resurgence and resummation, which have become a
subject of considerable interest in the study of gauge theories in recent
years (see \cite{Pasquetti:2009jg,Dunne:2013ada}, for example, \cite%
{Russo:2012kj,Honda:2016vmv,Dorigoni:2017smz} for supersymmetric gauge
theories and localization and \cite{Aniceto:2018bis} for a review).

We advance that the issue of Stokes lines crossing will not appear while
moving the physical parameters (which is essentially here the FI parameter)
of a given theory, unless we take $\eta \rightarrow \infty $. However, for
large but finite $\eta $, we will see that the crossing occurs when the
number of flavours is smaller than a certain value, in which case the theory
becomes a \textit{bad} theory.

The paper is organized as follows. In the next Section, we show how (\ref%
{mod1}) and (\ref{s-mod1}) can be evaluated by mapping the model into a
random matrix ensemble and using the results in \cite{FK}, which contains,
among other results, a new extension of the Selberg integral. We show that
the analytical continuation of the Beta function (\ref{Euler}) extends the
results to any number of flavours and to a non-zero FI parameter $\eta $. We
use this also to obtain an analytical expression for the Wilson loop. Four
specific sets of Wilson loops with FI parameter are studied in detail:
symmetric, antisymmetric, rectangular and hook representations, obtaining
explicit expressions in terms of Gamma functions of complex argument.

In Section 3, we focus exclusively on the partition function and show that
the matrix model (\ref{mod1}) and its counterparts for the orthogonal and
symplectic gauge groups, are all easily mapped into the Selberg integral,
which gives an exact evaluation of the free energies in terms of $G$-Barnes
functions. We also study the asymptotics for $\eta =0$ of the $U(N)$ free
energy for large $N$ and constant Veneziano parameter $\zeta =N_{f}/N$,
showing that the leading term is $f\left( \zeta \right) N^{2}$ instead of $%
N^{2}\log N$ and we determine $f\left( \zeta \right) $ and the subleading
contributions to the free energy. An integral representation of the
Mellin-Barnes type is given for the case of a $SU(N)$ gauge group and the $%
SU(2)$ case is computed explicitly in two ways.

In the last Section we study the asymptotics of the free energy, with FI
parameter, discussing the crossing of Stokes lines and the ensuing
appearance of exponentially small contributions, coming from expansions of
Gamma functions of complex argument, in the asymptotic expansion of the
partition functions. Explicit duality tests are carried out for the
analytical expressions obtained for the partition functions. We conclude
with some open directions for further work.

\section{Exact evaluation of the Wilson loop average}

We want the analytical evaluation (\ref{s-mod1}). For this, notice that the
usual change of variables $e^{\mu }=y$,\footnote{%
Precisely, we use $e^{\mu }=y$ and then $ye^{m}=x$.} useful when matrix
model contains the hyperbolic version of the Vandermonde determinant,
immediately leads to the following matrix model%
\begin{eqnarray}
\left\langle W_{\lambda }(N)\right\rangle &=&\frac{1}{Z}\int_{\left[
0,\infty \right) ^{N}}{d^{N}\!y}\prod_{i=1}^{N}\frac{e^{\frac{mN_{f}}{2}%
}y_{i}^{i\eta +\frac{N_{f}}{2}-N}\ }{(1+e^{m}y_{i})^{N_{f}}}s_{\lambda
}(y_{1},...,y_{N})\prod_{i<j}(y_{i}-y_{j})^{2}  \label{mapped} \\
&=&\frac{e^{-mN\left( i\eta +\frac{\left\vert \lambda \right\vert }{N}%
\right) }}{Z}\int_{\left[ 0,\infty \right) ^{N}}{d^{N}\!x}\prod_{i=1}^{N}%
\frac{x_{i}^{i\eta +\frac{N_{f}}{2}-N}\ }{(1+x_{i})^{N_{f}}}s_{\lambda
}(x_{1},...,x_{N})\prod_{i<j}(x_{i}-x_{j})^{2}.  \notag
\end{eqnarray}%
The mass dependence is accounted for in the prefactor%
\begin{equation*}
\exp \left[ -mN\left( i\eta +\frac{\left\vert \lambda \right\vert }{N}%
\right) \right] ,
\end{equation*}%
and absent in the matrix model itself, as can be also simply seen directly
from (\ref{mod1}). The same holds for the normalizing partition function
term $Z$, since it has the same matrix model representation (\ref{mapped})
but without the Schur polynomial. Therefore, cancelling the common
mass-dependent prefactor, overall, only the term $\exp \left( -m\left\vert
\lambda \right\vert \right) $ will remain.

In contrast to the case of Chern-Simons theory, the absence of a Gaussian
factor in the matrix model representation (\ref{mod1}) implies that the $%
x^{i\eta +N_{f}/2-N}$ factor in the weight function is not removed with a
shift of the above change of variables, such as in \cite{Tierz}. Without
such term the integrand in (\ref{mapped}) is known to represent the joint
probability distribution function of an ensemble of complex matrices \cite%
{Forrain}. The model can then be solved by using for example\footnote{%
This leads to a discussion of the FI term, as this argument seemingly
requires that the FI is either taken to be $0$ (as is the case in \cite%
{Grassi:2014vwa}), or it is such that $i\eta \in 
%TCIMACRO{\U{2124} }%
%BeginExpansion
\mathbb{Z}
%EndExpansion
$. The latter choice has been discussed in order to avoid restriction on the
charge (size of the partition) of the Wilson loop average \cite%
{Assel:2015oxa} (where the choice is implemented through a modification of
the contour integration).} $\prod_{i=1}^{N}x_{i}^{n}s_{\lambda
}(x_{1},...,x_{N})=s_{\lambda +(n^{N})}(x_{1},...,x_{N})$, together with the
exact results in \cite[Lema 3, (a)]{FK}. However, directly employing the
later work \cite{FK2} instead, gives an immediate solution.

The implication of this solution for the gauge theories is what we discuss
here. The central result in \cite{FK2}, for us, is the following: Let $k$
and $n$ be nonnegative integers\footnote{%
We slightly change their notation, since their $m$ parameter could be
confussed with the mass here.}, then for any partition $\lambda $ such that $%
l(\lambda )\leq k$ and $l(\lambda ^{^{\prime }})\leq n$%
\begin{equation}
\frac{s_{\lambda }(1_{n})s_{\lambda }(1_{k})}{s_{\lambda ^{\prime }}(1_{a})}=%
\frac{1}{C_{n,k}^{a}}\int\limits_{0}^{\infty }\hspace{-0.7ex}\ldots \hspace{%
-0.7ex}\int\limits_{0}^{\infty }s_{\lambda }(t_{1},\ldots t_{k})\Delta
^{2}(t_{1},\dots ,t_{k})\prod_{j=1}^{k}\frac{t_{j}^{n-k}dt_{j}}{%
(1+t_{j})^{a+n+k}},  \label{FKeq}
\end{equation}%
with%
\begin{equation}
C_{n,k}^{a}=\int\limits_{0}^{\infty }\hspace{-0.7ex}\ldots \hspace{-0.7ex}%
\int\limits_{0}^{\infty }\Delta ^{2}(t_{1},\dots ,t_{k})\prod_{j=1}^{k}\frac{%
t_{j}^{n-k}dt_{j}}{(1+t_{j})^{a+n+k}}=\prod_{j=0}^{k-1}\frac{\Gamma \left(
j\right) \Gamma (j+n-k+1)\Gamma (a+j+1)}{\Gamma (a+n+j+1)}.  \label{ZSelb}
\end{equation}%
The later expression follows from Selberg's integral \cite[Lema 3, (a)]%
{FK,FK2} whereas (\ref{FKeq}) is a novel extension of the Selberg integral,
obtained in \cite{FK2} and previously, in a simpler form, in \cite[Lema 3,
(a)]{FK}. Notice that the double constraint above on the partition $\lambda $
of the Schur polynomial, necessarily implies that\footnote{%
Recall that $s_{\lambda }(t_{1},\ldots t_{m})=0$ if $l(\lambda )>m$.} $%
N_{f}>2N$ which is also the regime identified in \cite{Assel:2015oxa}.
Indeed, for these values the integrals are manifestly convergent. This
regime corresponds to the \textit{good} or \textit{ugly} classification in 
\cite{Gaiotto:2008ak}. In the last Section, while studying the asymptotics
of the observables, we will be naturally lead also into considering the
setting $N<N_{f}<2N$, which corresponds to \textit{bad} theories.

As we see from the expressions below, involving Schur polynomials, in
principle it seems we need to consider the case of an even number of
flavours and take, at least for the moment the restricted view, above
exposed, for the FI parameter. For definiteness, we take now $\eta =0$ but
then below, we show how to lift this restriction, using the analytical
continuation given by the beta function (\ref{Euler}). The Wilson loop is
then%
\begin{equation}
\left\langle W_{\lambda }(N)\right\rangle =e^{-m\left\vert \lambda
\right\vert }\frac{s_{\lambda }(1_{N_{f}/2})s_{\lambda }(1_{N})}{s_{\lambda
^{\prime }}(1_{N_{f}/2-N})},  \label{Wl}
\end{equation}%
which, by using the Weyl dimension formula for the specialization of the
Schur polynomials%
\begin{equation}
s_{\mu }(1^{N})=\frac{1}{G(N+1)}\prod_{1\leq j<k\leq N}(\mu _{j}-\mu
_{k}+k-j),
\end{equation}%
valid when $l(\mu )\leq N$, can also be written as%
\begin{equation}
\left\langle W_{\lambda }(N)\right\rangle =\frac{e^{-m\left\vert \lambda
\right\vert }G(N_{F}/2-N+1)}{G(N+1)G(N_{F}/2+1)}\frac{\prod_{1\leq j<k\leq
N}(\lambda _{j}-\lambda _{k}+k-j)\prod_{1\leq j<k\leq N_{F}/2}(\lambda
_{j}-\lambda _{k}+k-j)}{\prod_{1\leq j<k\leq _{N_{F}/2-N}}(\lambda
_{j}^{\prime }-\lambda _{k}^{\prime }+k-j)}  \label{Wl2}
\end{equation}%
and the partition function, from (\ref{ZSelb}) is%
\begin{equation}
Z_{N}=e^{-imN\eta }\frac{G\left( N+2\right) G(\frac{N_{f}}{2}-i\eta +1)G(%
\frac{N_{f}}{2}+i\eta +1)G(N_{f}-N+1)}{G(\frac{N_{f}}{2}-N-i\eta +1)G(\frac{%
N_{f}}{2}-N+i\eta +1)G(N_{f}+1)}.  \label{Zresult}
\end{equation}%
Looking for potential poles or zeroes of the expressions, recall that the
G-Barnes function is an entire function and its zeroes are located at $%
G(-n)=0$ for $n=0$ and $n\in 
%TCIMACRO{\U{2115} }%
%BeginExpansion
\mathbb{N}
%EndExpansion
$. As expected, there is a drastic difference with regards to convergence,
according to the FI parameter. For $\eta =0,$ the two G-Barnes in the
denominator do give zeroes, precisely for $N_{f}<2N-1$ and we have the
known, expected, divergence of the partition function in this case. This is
the well-known divergence of the partition function \cite{Yaakov:2013fza}.

For $\eta \neq 0,$ there is no divergence of the partition function, since
the G-Barnes of the denominators do not have a zero there, for a bad theory
with $N_{f}\leq 2(N-1)$. The case of complex $\eta $ \cite{Closset:2012vg}
will be briefly touched upon in the last Section. Notice that the G-Barnes
function part in (\ref{Zresult}) is symmetric under $\eta \rightarrow -\eta $%
. Therefore, due to the imaginary prefactor $Z_{N}(-\eta )=\overline{%
Z_{N}(\eta )}$, as also happens when there is a Chern-Simons term \cite%
{Closset:2012vg,Giasemidis:2015ial}. Note also that (\ref{Zresult}) admits
alternative equivalent expressions, for example involving Gamma functions.
This will be relevant below when discussing the $SU(N)$ case and also at the
end, when studying duality.

Regarding Wilson loops it seems that the required specialization of Schur
polynomials puts some restriction on the parameters of our model but
actually these admit an expression in terms of Beta functions and this
provides an extension to the whole complex plane. We show this explicitly
now.

\subsection{Symmetric and antisymmetric representations}

Thus, we focus now on two interesting specific instances of (\ref{Wl}),
namely antisymmetric representation and symmetric representation. Recall how
difficult these specific cases are to analyze in $\mathcal{N}=4$ $U(N)$ SYM
theory in four dimensions, even though the corresponding matrix model there
is a Gaussian ensemble (see \cite{Zarembo:2016bbk} for a recent review).

In contrast, in this case, the solvability of the model, which is in general
that of a multidimensional beta function (Selberg integral), reduces to that
of the ordinary beta integral. This leads to compact exact expressions valid
for all $N$. Precisely, recalling the Euler integral identity (\ref{Euler}),
then the elementary and homogeneous symmetric polynomials are given in terms
of the Beta function%
\begin{equation}
\frac{1}{e_{r}(1_{n})}=(n+1)\int_{0}^{\infty }\frac{t^{r}}{(1+t)^{n+2}}dt%
\text{ \ \ \ and \ \ }\frac{1}{h_{r}(1_{n})}=(n-1)%
\int_{0}^{1}t^{r}(1-t)^{n-2}dt.  \label{eh}
\end{equation}%
Then, with $\lambda $ a partition of length $1$, $\lambda =(r)$, we have that%
\begin{equation}
s_{\lambda }(1_{n})=h_{r}(1_{n})=\binom{n+r-1}{r}\text{ \ \ \ and \ \ }%
s_{\lambda ^{\prime }}(1_{n})=e_{r}(1_{n})=\binom{n}{r}.  \label{Gamma}
\end{equation}%
The general Wilson loop expression (\ref{Wl}) adopts a very concrete
expression in terms of Gamma functions, for a symmetric representation%
\begin{equation*}
\left\langle W_{(r)}(N)\right\rangle =e^{-mNr}\frac{\Gamma \left( \frac{N_{f}%
}{2}+r\right) \Gamma \left( N+r\right) \Gamma \left( \frac{N_{f}}{2}%
-N-r+1\right) }{\Gamma \left( r+1\right) \Gamma \left( \frac{N_{f}}{2}%
\right) \Gamma \left( N\right) \Gamma \left( \frac{N_{f}}{2}-N+1\right) },
\end{equation*}%
and the antisymmetric representation%
\begin{equation*}
\left\langle W_{(1)^{r}}(N)\right\rangle =\frac{e^{-mNr}\Gamma \left( \frac{%
N_{f}}{2}+1\right) \Gamma \left( N+1\right) \Gamma \left( \frac{N_{f}}{2}%
-N\right) }{\Gamma \left( r+1\right) \Gamma \left( \frac{N_{f}}{2}%
-r+1\right) \Gamma (N-r+1)\Gamma (\frac{N_{f}}{2}-N+r)}.
\end{equation*}%
Notice that these expressions do not restrict $N_{f}$ to be even. Likewise,
the expressions in terms of Gamma functions suggest the consideration of the
FI parameter. In that case, we would have%
\begin{equation}
\left\langle W_{\lambda }(N,\eta )\right\rangle =e^{-m\left\vert \lambda
\right\vert }\frac{s_{\lambda }(\boldsymbol{1}_{\frac{N_{f}}{2}+i\eta
})s_{\lambda }(\boldsymbol{1}_{N})}{s_{\lambda ^{\prime }}(\boldsymbol{1}_{%
\frac{N_{f}}{2}-N-i\eta })}.  \label{Wfact}
\end{equation}%
Because of (\ref{eh}), each term in this expression is a Beta function, two
of these have one of their two parameters complex, so one can use (\ref%
{Euler})\footnote{%
A way to prove (\ref{Euler}) is by induction when $p$ is an integer, and
since the integral is bounded and analytical for $\func{Re}p,$ $\func{Re}%
q\geq 0$ -and so is the r.h.s. of the formula- then by Carlson theorem \cite%
{Carlson}, the expression follows for complex $p$ and $q$. However, to prove
uniqueness of the analytical extension of the Wilson loops, one needs to
guarantee also that the factorized expression (\ref{Wfact}) also holds for
complex values. Here, we just extended each piece in a unique way, by using
the Beta function.} and the resulting expressions with a generic $N_{f}$ and
a non-zero FI parameter are then given by%
\begin{eqnarray}
\left\langle W_{(r)}(N,\eta )\right\rangle &=&e^{-mr}\frac{\Gamma \left(
i\eta +\frac{N_{f}}{2}+r\right) \Gamma \left( N+r\right) \Gamma \left( \frac{%
N_{f}}{2}-N-i\eta -r+1\right) }{\Gamma \left( r+1\right) \Gamma \left( \frac{%
N_{f}}{2}+i\eta \right) \Gamma \left( N\right) \Gamma \left( \frac{N_{f}}{2}%
-N-i\eta +1\right) },  \label{Wr} \\
\left\langle W_{(1)^{r}}(N,\eta )\right\rangle &=&\frac{e^{-mr}\Gamma \left( 
\frac{N_{f}}{2}+i\eta +1\right) \Gamma \left( N+1\right) \Gamma \left( \frac{%
N_{f}}{2}-N-i\eta \right) }{\Gamma \left( r+1\right) \Gamma \left( \frac{%
N_{f}}{2}-r+i\eta +1\right) \Gamma (N-r+1)\Gamma (\frac{N_{f}}{2}-N-i\eta +r)%
}.  \label{W1r}
\end{eqnarray}%
Note that in these expressions the Gamma functions can be traded for
Pochhammer symbols, using $r-1$ times $\Gamma (z+1)=z\Gamma (z)$. Some
simple checks can be quickly made. For example for $r=0$ the expressions
above give $1$, as expected and for $r=1$ (fundamental representation) both
expressions coincide, giving 
\begin{equation*}
\left\langle W_{(1)}(N,\eta )\right\rangle =e^{-m}\frac{N\Gamma \left( i\eta
+\frac{N_{f}}{2}+1\right) \Gamma \left( \frac{N_{f}}{2}-N-i\eta \right) }{%
\Gamma \left( \frac{N_{f}}{2}+i\eta \right) \Gamma \left( \frac{N_{f}}{2}%
-N-i\eta +1\right) }=Ne^{-m}\frac{i\eta +\frac{N_{f}}{2}}{\frac{N_{f}}{2}%
-N-i\eta }.
\end{equation*}%
Likewise, as for the partition function (but this time due to the Gamma
functions and not the exponential prefactor, which is real), we have that $%
\left\langle W_{\mu }(N,-\eta )\right\rangle =\overline{\left\langle W_{\mu
}(N,\eta )\right\rangle }$.

The case of antisymmetric representation with $N$ boxes can be seen as a
partition function computation because $%
s_{(1^{N})}(x_{1},...,x_{N})=e_{N}(x_{1},...,x_{N})=$ $\tprod%
\nolimits_{i=1}^{N}x_{i}$, see below for this point of view in the more
general case of a rectangular representation. Here, using (\ref{W1r}) we
have that%
\begin{equation}
\left\langle W_{(1^{N})}(N,\eta )\right\rangle =e^{-mN}\frac{\Gamma \left( 
\frac{N_{f}}{2}+i\eta +1\right) \Gamma \left( \frac{N_{f}}{2}-N-i\eta
\right) }{\Gamma \left( \frac{N_{f}}{2}-N+i\eta +1\right) \Gamma \left( 
\frac{N_{f}}{2}-i\eta \right) }.  \label{W1N}
\end{equation}%
Notice that it is as the result for the fundamental but with the complex
conjugate in the denominator.

\subsection{Rectangular partitions}

There are more general representations that also can be studied very
explicitly. In particular, the case of rectangular partitions is specially
interesting. Recall the statement made above, before (\ref{FKeq}), on
rectangular partitions: if we consider the partition of length $N$, $\left(
l,l,...,l\right) $ which we denote by $l^{N}=(l^{N})$ then, assuming that $%
\lambda $ is a partition of length equal or lower than $N$, we have that%
\begin{equation}
s_{\lambda +l^{N}}(x_{1},...,x_{N})=e_{N}^{l}(x_{1},...,x_{N})s_{\lambda
}(x_{1},...,x_{N})\text{,}  \label{ext_rect}
\end{equation}%
which follows by recalling that $e_{N}(x_{1},...,x_{N})=\tprod%
\nolimits_{i=1}^{N}x_{i}$. Thus, as a simple extension of the result above
for the $\left( 1^{N}\right) $ representation, the case where the
representation is described by a rectangular partition (with number of rows
equal to the rank $N$ of the gauge group), the matrix model giving the
Wilson loop has the same form as the one for the partition function, with a
shift of parameters. Thus,%
\begin{equation*}
\left\langle W_{(l^{N})}(N,N_{f},\eta )\right\rangle =\frac{Z_{N}\left(
N,N_{f},\eta -il\right) }{Z_{N}\left( N,N_{f},\eta \right) }=e^{-mNl}\frac{%
\prod\limits_{j=1}^{l}\Gamma \left( \frac{N_{f}}{2}+i\eta +1+l-j\right)
\Gamma \left( \frac{N_{f}}{2}-N-i\eta +1-j\right) }{\prod\limits_{j=1}^{l}%
\Gamma \left( \frac{N_{f}}{2}-i\eta +1-j\right) \Gamma \left( \frac{N_{f}}{2}%
-N+i\eta +l+1-j\right) },
\end{equation*}%
where, in addition to taking the quotient using (\ref{Zresult}), we have
also iteratively applied $G(z+1)=\Gamma (z)G(z)$, which also will be crucial
below to check Seiberg duality. Notice that for $l=1$, it coincides with (%
\ref{W1N}), as it should.

Likewise, from (\ref{ext_rect}) it follows that%
\begin{equation*}
\left\langle W_{\lambda +l^{N}}(N,N_{f},\eta )\right\rangle =\left\langle
W_{\lambda }(N,N_{f},\eta -il)\right\rangle
\end{equation*}%
Thus, this case is equivalent to that of a partition function with a complex
FI\ parameter. This setting will be briefly discussed again at the end, when
considering the asymptotics of $G$-Barnes functions and the crossing of
Stokes lines.

\subsection{Hook representations}

The last explicit case that we analyze is the one corresponding to
partitions represented by hooks, $\lambda =(r-s,1^{s})$. As above $%
\left\vert \lambda \right\vert =r$. Notice that this notation describes a
Young tableaux of a row of size $r-s$ and a column of size $s+1$ as $1^{s}$
contains $s$ boxes below the upper-left box of the tableaux. Therefore, $%
\lambda ^{\prime }=(s+1,1^{r-s-1})$ and we have%
\begin{equation}
s_{\lambda }(1_{n})=\frac{\Gamma \left( n+r-s\right) }{r\Gamma (r-s)\Gamma
(s+1)\Gamma (n-s)}\text{ \ and }s_{\lambda ^{\prime }}(1_{n})=\frac{\Gamma
\left( n+s+1\right) }{r\Gamma (s+1)\Gamma (r-s)\Gamma (n-r+s+1)},
\label{hookdims}
\end{equation}%
notice that for $s=0$, then $\lambda =(r)$ and $\lambda ^{\prime
}=(1,1^{r-1})=(1^{r})$ and the above expressions reduce to the ones for the
homogeneous and elementary symmetric polynomials, respectively (\ref{Gamma})
as it should. Likewise, the same consistency check is done for the dual
situation, given by $s=r-1$. Then, again using (\ref{Wfact}), we obtain%
\begin{equation}
\left\langle W_{(r-s,1^{s})}(N,\eta )\right\rangle =\frac{e^{-mr}\Gamma
\left( \frac{N_{f}}{2}+i\eta +r-s\right) \Gamma \left( N+r-s\right) \Gamma (%
\frac{N_{f}}{2}-N-i\eta -r+s+1)}{r\Gamma (r-s)\Gamma (s+1)\Gamma (\frac{N_{f}%
}{2}+i\eta -s)\Gamma (N-s)\Gamma \left( \frac{N_{f}}{2}-N-i\eta +s+1\right) }%
,  \label{Whook}
\end{equation}%
if $s=0$ this expression indeed reduces to (\ref{Wr}) and if $s=r-1$ it then
gives (\ref{W1r}). Notice that, alternatively to (\ref{Wfact}), one could
also use (\ref{Whook}), together with Giambelli determinant formula, to
obtain a Wilson loop in other representations, out of the hook expressions.

The asymptotics of these Wilson loop expressions is particularly rich
because having complex arguments in the Gamma functions, then the crossing
of Stokes lines (and related phenomena, like Berry smoothening transitions
across lines \cite{Berry}, etc.) appears. We will discuss this at the end
focussing more on the free energy.

\section{$O(2N),O(2N+1)$ and $Sp(2N)$ cases, $U(N)$ asymptotics and integral
representation for $SU(N)$}

In this Section, we show that the partition function of the matrix models
also follows, after some change of variables, from the famous evaluation of
the Selberg integral \cite{For}%
\begin{eqnarray}
S_{N}(\lambda _{1},\lambda _{2},\gamma ) &:&=\int_{0}^{1}\cdots
\int_{0}^{1}\,\prod_{i=1}^{N}t_{i}^{\lambda _{1}}(1-t_{i})^{\lambda
_{2}}dt_{i}\prod_{1\leq i<j\leq n}\left\vert t_{i}-t_{j}\right\vert
^{2\gamma }  \label{SelInt} \\
&=&\prod_{j=0}^{N-1}\frac{\Gamma (\lambda _{1}+1+j\gamma )\Gamma (\lambda
_{2}+1+j\gamma )\Gamma (1+(j+1)\gamma )}{\Gamma (\lambda _{1}+\lambda
_{2}+2+(N+j-1)\gamma )\Gamma (1+\gamma )}.  \notag
\end{eqnarray}%
The evaluation of this integral is valid for complex parameters $\lambda
_{1},\lambda _{2},\gamma $ such that%
\begin{equation}
\Re (\lambda _{1})>0,~\Re (\lambda _{2})>0,~\Re (\gamma )>-\min \{1/N,\Re
(\lambda _{1})/(N-1),\Re (\lambda _{2})/(N-1)\},  \label{SelIntpv}
\end{equation}%
corresponding to the domain of convergence of the integral. This evaluation
is useful not only for the $U(N)$ case above discussed, but also when the
gauge group is the symplectic or the orthogonal group. These latter cases
were studied with the Fermi gas formalism in \cite{Mezei:2013gqa}. We
conclude this Section with a discussion on how to obtain the $SU(N)$ case by
integration of the $U(N)$ result (\ref{Zresult}) over the FI parameter.

\subsection{$U(N)$ case revisited and free energy behavior for large $N$}

Notice that, with only a very minor modification of a change of variables
proposed in \cite[Ex. 4.1.3]{For}%
\begin{equation*}
t_{l}=\frac{1}{e^{\left( s_{l}+m\right) /2}+1},
\end{equation*}%
the Selberg integral can be written as%
\begin{eqnarray*}
S_{N}(\lambda _{1},\lambda _{2},\lambda ) &=&e^{-\left( \lambda _{1}-\lambda
_{2}\right) mN/2}\int_{-\infty }^{\infty }\cdots \int_{-\infty }^{\infty
}\prod_{i=1}^{N}\frac{e^{-\left( \lambda _{1}-\lambda _{2}\right)
s_{i}/2}ds_{i}}{\left( 2\cosh (\frac{1}{2}(s_{i}+m))\right) ^{\lambda
_{1}+\lambda _{2}+2+2\lambda (N-1)}} \\
&&\times \prod_{i<j}\left( 2\sinh (\frac{1}{2}(s_{i}-s_{j}))\right)
^{2\lambda },
\end{eqnarray*}%
and hence, choosing $\lambda =1,$ $\lambda _{1}=\left( 2\mathrm{i}\eta
-2N+N_{f}\right) /2$ and $\lambda _{2}=\left( N_{f}-2N-2\mathrm{i}\eta
\right) /2$, we have that%
\begin{eqnarray}
Z_{N} &=&e^{-i\eta mN}S_{N}\left( 1,\frac{2\mathrm{i}\eta -2N+N_{f}}{2},%
\frac{N_{f}-2N-2\mathrm{i}\eta }{2}\right)  \notag \\
&=&e^{-i\eta mN}\prod_{j=0}^{N-1}\frac{\Gamma (\mathrm{i}\eta
+1+N_{f}/2-N+j)\Gamma (-\mathrm{i}\eta +1+N_{f}/2-N+j)\Gamma (j+2)}{\Gamma
(-N+N_{f}+1+j)}.  \label{Z2}
\end{eqnarray}%
\ \ Once again, the partition function can be written in terms of Barnes $G$%
-functions%
\begin{equation}
Z_{N}=e^{-i\eta mN}\frac{G(\mathrm{i}\eta +1+N_{f}/2)G(-\mathrm{i}\eta
+1+N_{f}/2)G(-N+N_{f}+1)G(N+2)}{G(\mathrm{i}\eta +1+N_{f}/2-N)G(-\mathrm{i}%
\eta +1+N_{f}/2-N)G(N_{f}+1)G(2)},  \label{ZG}
\end{equation}%
which, as expected, coincides with (\ref{Zresult}). The expression
simplifies when $\eta =0$, giving%
\begin{equation}
Z_{N_{f}}^{U(N)}(\eta =0)=\frac{G^{2}(1+N_{f}/2)G(-N+N_{f}+1)G(N+2)}{%
G^{2}(1+N_{f}/2-N)G(N_{f}+1)}.  \label{ZG0}
\end{equation}%
In the $\eta =0$ case, as seen also from the condition (\ref{SelIntpv}), the
result holds for $N_{f}\geq 2N-1$, a well-known result \cite{Yaakov:2013fza}%
. In the large $N$ and $N_{f}$ limit, therefore, this implies a lower bound
for the Veneziano parameter $\zeta =N_{f}/N>2$. Now, one can immediately
study the large $N_{f}\rightarrow \infty $ and $N\rightarrow \infty $
behavior in (\ref{ZG0}), directly using the original expansion by Barnes%
\begin{equation}
\log G(z+1)=\frac{1}{2}-\log A+\frac{z}{2}\log 2\pi +\left( \frac{z^{2}}{2}-%
\frac{1}{12}\right) \log z-\frac{3z^{2}}{4}+\sum_{k=1}^{N}\frac{\mathrm{B}%
_{2k+2}}{4k(k+1)z^{2k}}+O\left( \frac{1}{z^{2N+2}}\right) ,  \label{Basympt}
\end{equation}%
where the constant $A$ is the Glaisher--Kinkelin constant. The solution can
be given in terms of $N$ and the Veneziano parameter, in which case the
partition function reads%
\begin{equation}
Z(N,\zeta ,\eta )=\frac{G(1+i\eta +\frac{\zeta N}{2})G(1-i\eta +\frac{\zeta N%
}{2})G(\left( \zeta -1\right) N+1)G(N+2)}{G(i\eta +\left( \zeta /2-1\right)
N+1)G(-i\eta +\left( \zeta /2-1\right) N+1)G(\zeta N+1)}.  \label{GVen}
\end{equation}%
For $\eta =0$ we can also for example study the first convergent case, given
by $N_{f}=2N+1$ or, more generally, $N_{f}=2N+k$ with $k\geq 1$ and finite,
which gives 
\begin{equation*}
Z_{2N+1}^{U(N)}(\eta =0)=\frac{G^{2}(1+N+\frac{k}{2})G(N+k+1)G(N+2)}{%
G^{2}(1+k/2)G(2N+1+k)},
\end{equation*}%
this covers an arbitrary large number of cases (indexed by $k$), and the
Veneziano parameter is always $2$. Now, for $\eta =0$ and taking into
account the asymptotics of the Barnes G-function, it follows that, for the
double scaling limit $N\rightarrow \infty $ and $N_{f}\rightarrow \infty $
with $\zeta =$ $N_{f}/N=\mathrm{cte}$ , if we write the free energy as $%
F_{N}\equiv \ln Z_{N}$, what would be the leading term in the free energy%
\begin{equation}
\frac{N^{2}}{2}\log N,  \label{preleading}
\end{equation}%
always cancels out and it is not present. This holds because, if we denote
by $z_{i}$ the terms $1+z_{i}$ in the arguments of the different G-Barnes in
(\ref{GVen}) with $\eta =0$\footnote{%
From $i=1$ to $4$ it denotes the arguments in the numerator and from $5$ to $%
7$ the ones in the denominator.}, then it holds that%
\begin{equation}
\sum\limits_{i=1}^{4}z_{i}^{2}-\sum\limits_{i=5}^{7}z_{i}^{2}=\sum%
\limits_{i=1}^{4}z_{i}-\sum\limits_{i=5}^{7}z_{i}=2N+1.  \label{leading}
\end{equation}%
The final term that comes from the $-3z^{2}/4$ piece in (\ref{Basympt})
cancels completely due to (\ref{leading}) and the same happens, partially,
for the term that results from the $(z^{2}/2)\log z$ piece in (\ref{Basympt}%
). The (\ref{preleading}) behavior that would follow from this piece
cancels, and it only remains a $N^{2}$ leading term that is multiplied by a
Veneziano parameter dependent factor. All together we have, for $%
N\rightarrow \infty $%
\begin{equation*}
F_{N}\sim N^{2}f\left( \zeta \right) +N\log N+N\left( \log 2\pi -\frac{3}{2}%
\right) +\frac{5}{12}\log N,
\end{equation*}%
where%
\begin{equation}
f\left( \zeta \right) =\frac{\zeta ^{2}}{2}\log \frac{\zeta }{2}+\left(
\zeta -1\right) ^{2}\log \left( \zeta -1\right) -2\left( \frac{\zeta }{2}%
-1\right) ^{2}\log \left( \frac{\zeta }{2}-1\right) -\zeta ^{2}\log \zeta .
\label{f-Ven}
\end{equation}%
Note that for the particular case of $\zeta =2$ all the terms in (\ref{f-Ven}%
) cancel, with the exception of the last one and then $f\left( \zeta \right)
=-4\log 2$. In general, while naively at first it may seem that $f\left(
\zeta \right) $ vanishes for large $\zeta $, it turns out that it is well
approximated by $f\left( \zeta \right) \simeq -2\zeta \log 2$ for large $%
\zeta $.

Recall that the free energy of the larger family of 3d $T_{\rho }^{\rho
^{\prime }}[SU(N)]$ theories, at leading order, has been studied in \cite%
{Assel:2012cp,Nishioka:2011dq} and, more recently, from the point of view of
gravitational duals, in \cite{Lozano:2016wrs} where it is shown that the
leading term is of the form (\ref{preleading}). This property also holds for
the simpler $T[SU(N)]$ theory, namely the $SU(N)$ theory with $N_{f}$
flavours of different masses $m_{l}$ with $l=1,...,N_{f}$ \cite%
{Assel:2012cp,Nishioka:2011dq}, and we have just studied the case when all
masses are equal, which leads to the cancellation of such a leading term.
For a very recent discussion of other 3d SYM theories with similar behavior
to the one obtained here, see \cite{Assel:2018vtq}.

\subsection{$O(2N),O(2N+1)$ and $Sp(2N)$ cases}

Now, setting the mass and the FI parameter $\eta =m=0$, the explicit
expression of the free energy when the gauge group are the orthogonal and
symplectic groups can be obtained, again with the same Selberg integral,
just by considering another change of variables. Indeed, with the change of
variables \cite[Ex. 4.1.3]{For}%
\begin{equation*}
t_{l}=\frac{2}{\cosh \left( s_{l}/2\right) +1}
\end{equation*}%
we have that%
\begin{eqnarray}
S_{N}(\lambda _{1},\lambda _{2},\lambda ) &=&2^{-N}\int_{-\infty }^{\infty
}\cdots \int_{-\infty }^{\infty }\prod_{i=1}^{N}\frac{\left( \sinh
^{2}\left( s_{l}/2\right) \right) ^{\lambda _{2}+1/2}ds_{l}}{\left( \cosh
^{2}(s_{i}/2)\right) ^{\lambda _{1}+\lambda _{2}+2+2\lambda (N-1)}}  \notag
\\
&&\times \prod_{i<j}\left( \sinh (\frac{1}{2}(s_{i}-s_{j}))\sinh (\frac{1}{2}%
(s_{i}+s_{j}))\right) ^{2\lambda }.  \label{Shyp2}
\end{eqnarray}%
Recall that the Vandermonde determinant of the matrix model, in case of the
orthogonal and symplectic gauge groups is the corresponding hyperbolic
version of the Haar measure, explicitly given by: 
\begin{eqnarray*}
&&\Delta _{O(2N)}^{2}(e^{is})=\prod_{i<j}\left( 2\sinh (\frac{1}{2}%
(s_{i}-s_{j}))2\sinh (\frac{1}{2}(s_{i}+s_{j}))\right) ^{2}, \\
\Delta _{O(2N+1)}^{2}(e^{is}) &=&\prod_{i<j}\left( 2\sinh (\frac{1}{2}%
(s_{i}-s_{j}))2\sinh (\frac{1}{2}(s_{i}+s_{j}))\right)
^{2}\prod_{i=1}^{N}\sinh ^{2}\left( s_{l}/2\right) , \\
\Delta _{Sp(2N)}^{2}(e^{is}) &=&\prod_{i<j}\left( 2\sinh (\frac{1}{2}%
(s_{i}-s_{j}))2\sinh (\frac{1}{2}(s_{i}+s_{j}))\right)
^{2}\prod_{i=1}^{N}\sinh ^{2}\left( s_{l}\right) .
\end{eqnarray*}%
Thus, we find that, in terms of the $S_{N}$ given by (\ref{SelInt}), the
partition functions read:%
\begin{eqnarray*}
Z_{N_{f}}^{O(2N)} &=&2^{2N(N-(N_{f}-3)/2)}S_{N}((N_{f}+1)/2-2N,-1/2,1) \\
Z_{N_{f}}^{O(2N+1)} &=&2^{2N(N-(N_{f}-3)/2)}S_{N}((N_{f}-1)/2-2N,+1/2,1) \\
Z_{N_{f}}^{Sp(2N)} &=&S_{N}((N_{f}-3)/2-2N,+1/2,1),
\end{eqnarray*}%
where for the symplectic case we have used a half-angle formula for the $%
\sinh $ function in (\ref{Shyp2}). The partition functions in terms of the G
Barnes function are then given by%
\begin{eqnarray*}
Z_{N_{f}}^{O(2N)} &=&2^{2N(N-(N_{f}-3)/2)}\frac{G\left(
(N_{f}+1)/2-N+1\right) G(N+1/2)G(N+2)G\left( N_{f}/2-N\right) }{G\left(
(N_{f}+1)/2-2N+1\right) G(1/2)G\left( N_{f}/2\right) }, \\
Z_{N_{f}}^{O(2N+1)} &=&2^{2N(N-(N_{f}-3)/2)}\frac{G\left(
(N_{f}-1)/2-N+1\right) G(N+3/2)G(N+2)G\left( N_{f}/2-N+1\right) }{G\left(
N_{f}/2+1\right) G(3/2)G\left( N_{f}/2+1/2\right) }, \\
Z_{N_{f}}^{Sp(2N)} &=&\frac{G\left( (N_{f}-3)/2-N+1\right) G(N+3/2)G\left(
N_{f}/2-N\right) }{G\left( N_{f}/2\right) G\left( (N_{f}-3)/2-2N+1\right)
G(3/2)}.
\end{eqnarray*}

\subsection{Mellin-Barnes representation for $SU(N)$}

To conclude this Section, notice that the $SU(N)$ case could also be
computed by integrating the $U(N)$ result over the FI parameter (as this is
equivalent to introduce a Dirac delta in the matrix model). We start by
rewriting the expression (\ref{Zresult}) in terms of Gamma functions for the
part containing the FI parameter. That is%
\begin{equation}
Z_{N}=e^{-imN\eta }\dprod\limits_{j=1}^{N}\left\vert \Gamma \left( \frac{%
N_{f}}{2}-j+i\eta +1\right) \right\vert ^{2}\frac{G\left( N+2\right)
G(N_{f}-N+1)}{G(N_{f}+1)},  \label{ZGamma}
\end{equation}%
which follows immediately by repeatedly using $G(z+1)=\Gamma (z)G(z)$.
First, we analyze the $SU(2)$ case. In the massless case, we can directly
use first Barnes lemma%
\begin{equation*}
\frac{1}{2\pi i}\int_{-i\infty }^{i\infty }\Gamma \left( a+x\right) \Gamma
\left( b+x\right) \Gamma \left( c-x\right) \Gamma \left( d-x\right) dx=\frac{%
\Gamma \left( a+c\right) \Gamma \left( a+d\right) \Gamma \left( b+c\right)
\Gamma \left( b+d\right) }{\Gamma \left( a+b+c+d\right) },
\end{equation*}%
which is an extension of the beta integral and equivalent to Gauss summation
of the hypergeometric function. Then, normalizing by $2\pi $ to account for
the Dirac delta representation as a Fourier transform of the identity, we
have 
\begin{eqnarray*}
Z_{SU(2)} &=&\frac{G\left( 4\right) G(N_{f}-1)}{\left( 2\pi \right)
G(N_{f}+1)}\int_{-\infty }^{\infty }\left\vert \Gamma \left( \frac{N_{f}}{2}%
+i\eta \right) \right\vert ^{2}\left\vert \Gamma \left( \frac{N_{f}}{2}%
+i\eta -1\right) \right\vert ^{2}d\eta \\
&=&\frac{G\left( 4\right) G(N_{f}-1)\Gamma \left( N_{f}\right) \Gamma \left(
N_{f}-1\right) ^{2}\Gamma \left( N_{f}-2\right) }{G(N_{f}+1)\Gamma \left(
2N_{f}-2\right) }.
\end{eqnarray*}%
Notice that the expression diverges for $N_{f}=1$ and $N_{f}=2$ and indeed
these two cases correspond to \textit{bad} theories. Of course, for this
case the direct computation of the matrix model (\ref{mod1}) with the $%
\delta \left( x_{1}+x_{2}\right) $ insertion is direct as well, with the
resulting one dimensional integral, giving: 
\begin{equation}
\widetilde{Z}_{SU(2)}=\int_{-\infty }^{\infty }\frac{4\sinh ^{2}x}{\left(
2\cosh \left( x/2\right) \right) ^{2N_{f}}}dx,  \label{Z2direct}
\end{equation}%
and, using for example (\ref{Abelian}), one checks that indeed $\widetilde{Z}%
_{SU(2)}=$ $Z_{SU(2)}$. The massive version of (\ref{Z2direct}) is also
solvable (for example, using the results in \cite{Russo:2016ueu}) and can be
related to Wilson loops in a $U(1)$ theory with $N_{f}$ hypermultiplets of
mass $m$ and $N_{f}$ hypermultiplets of mass $-m$, but this will be
discussed elsewhere.

From the expression (\ref{ZGamma}), it follows that the partition function
for the $SU(N)$ theory admits a one-dimensional Mellin-Barnes type of
integral representation%
\begin{equation}
Z_{SU(N)}=\frac{G\left( N+2\right) G(N_{f}-N+1)}{G(N_{f}+1)}\int_{-\infty
}^{\infty }e^{-imN\eta }\dprod\limits_{j=1}^{N}\Gamma \left( \frac{N_{f}}{2}%
-j+i\eta +1\right) \Gamma \left( \frac{N_{f}}{2}-j-i\eta +1\right) d\eta ,
\label{MB}
\end{equation}%
with a Fourier kernel also, in the massive case. Both asymptotics and a full
analytical solution are possible but not entirely immediate (a formula for
generic values of all parameters ends up being quite involved), requiring an
analysis in itself, and hence it will be presented elsewhere.

\section{On exponential asymptotics and duality}

The presence of the FI parameter has an interesting implication: the
expressions for the free energies and the Wilson loops, given above, are in
terms of G-Barnes or Gamma functions of \textit{complex} arguments. This
leads naturally to consider the behavior of these functions in the whole
complex plane, where a richer behavior, involving Stokes lines \cite%
{Berry,Nemes}, is well-known to emerge.

Let us remind first that $G\left( {z+1}\right) $ and $\Gamma \left( z\right) 
$ (we will directly look at the asymptotic expansion of their logarithm) has
Stokes lines at $z=\pm \pi /2$. Thus, looking at the logarithm of the
solution (\ref{Zresult}), it appears that for certain values of the rank of
the gauge group $N$ and the number of flavours $N_{f}$ (and, in the case of
the Wilson loops, the size of the representation), we will have the
appearance of exponentially small contributions in the asymptotic expansions
of the observables. Notice that these are not phase transitions or
crossovers within eventual different regimes of the theory, since the
controlling parameters are number of flavours and colors.

We focus on the $U(N)$\ free energy, given by the logarithm of (\ref{ZG}).
Its analysis immediately follows from considering the mathematical results
on the G Barnes function \cite{Pasquetti:2009jg,Nemes}. The exponentially
improved asymptotic expansion of the G-Barnes function reads \cite{Nemes}%
\begin{equation}
\begin{split}
\log G\left( {z+1}\right) & \sim \frac{1}{4}z^{2}+z\log \Gamma \left( {z+1}%
\right) -\left( {\frac{1}{2}z\left( {z+1}\right) +\frac{1}{{12}}}\right)
\log z-\log A \\
& +\sum\limits_{k=1}^{\infty }{S_{k}\left( \theta \right) e^{\pm 2\pi ikz}}%
+\sum\limits_{n=1}^{\infty }{\frac{{B_{2n+2}}}{{2n\left( {2n+1}\right)
\left( {2n+2}\right) z^{2n}}}},
\end{split}
\label{Gasymp}
\end{equation}%
where%
\begin{equation}
S_{k}\left( \theta \right) =%
\begin{cases}
0 & \text{ if }\left\vert \theta \right\vert <\frac{\pi }{2} \\ 
\mp \cfrac{1}{2}\cfrac{1}{{2\pi ik^2 }} & \text{ if }\theta =\pm \frac{\pi }{%
2} \\ 
\mp \cfrac{1}{{2\pi ik^2 }} & \text{ if }\frac{\pi }{2}<\left\vert \theta
\right\vert <\pi ,%
\end{cases}
\label{eq25}
\end{equation}%
and $\theta =\arg z$. The upper or lower sign is taken according to $z$
being in the upper or lower half-plane. The term with $S_{k}\left( \theta
\right) $ describes the Stokes singularities and the rest of (\ref{Gasymp})
is the asymptotic expansion of the G Barnes function, as given in \cite{FL-G}%
. Note that this is not the original asymptotic expansion by Barnes, used
above, but, using the asymptotics for the $\log \Gamma \left( {z+1}\right) $
in (\ref{Gasymp}) (including exponentially small contributions) one can
write down the analogous result for the Barnes form above (\ref{Basympt}).
At any rate, as will be seen below, we crucially need both asymptotics for
our analysis of the free energy.

The four $G(1+z)$ functions of the free energy (taking the logarithm of (\ref%
{Zresult})) with complex argument, have a $z$ variable given by:%
\begin{eqnarray*}
z_{1,\pm } &=&\pm \mathrm{i}\eta +N_{f}/2-N, \\
z_{2,\pm } &=&\pm \mathrm{i}\eta +N_{f}/2.
\end{eqnarray*}%
The Stokes lines are located at $\pm \pi /2,$ hence for large but finite $%
\eta $, we need to look when the real part of the arguments becomes $0$
and/or negative. A null or negative real part argument can only physically
happen for the $z_{1,\pm }$ case above. The first case, which is right at
the Stokes line, corresponds to $N_{f}=2N$ which is a self-dual case in
terms of dualities (see below). The first actual crossing of a Stokes line
occurs for $N_{f}=2N-1$, the ugly case, and the rest is then for 
\begin{equation*}
N_{f}<2N-1,
\end{equation*}%
which is well-known to correspond to the so-called \textit{bad} theories.
Therefore, we discuss below this eventual crossing of the Stokes line also
in the context of the analysis of \textit{bad}, \textit{good} and \textit{%
ugly} theories \cite{Assel:2017jgo,Yaakov:2013fza} and dualities below.

We need the asymptotics of the Gamma function too, whose Stokes phenomena is
similar to that of the G-Barnes function above, with the Stokes lines also
at $\pm \pi /2$. One difference is the different decay of the $S_{k}\left(
\theta \right) ${\ }coefficients (Stokes multipliers), where the quadratic
decay in the G-Barnes is a linear decay in the Gamma function case \cite%
{Nemes2}. More crucially, there is a sign difference in the respective
Stokes multipliers, as we show in what follows. For the Gamma function the
following asymptotic expansion holds as $z\rightarrow \infty $%
\begin{equation*}
\log \Gamma ^{\ast }(z)\sim \sum_{n=1}^{\infty }\frac{B_{2n}}{%
2n(2n-1)z^{2n-1}}-%
\begin{cases}
0 & \text{ if }\left\vert \theta \right\vert <\frac{\pi }{2} \\ 
\frac{1}{2}\log (1-e^{\pm 2\pi iz}) & \text{ if }\theta =\pm \frac{\pi }{2}
\\ 
\log (1-e^{\pm 2\pi iz}) & \text{ if }\frac{\pi }{2}<\left\vert \theta
\right\vert <\pi ,%
\end{cases}%
.
\end{equation*}%
The expansion of the logarithm brings the asymptotics in the same form as
above%
\begin{equation}
\log \Gamma ^{\ast }(z)\sim \sum_{n=1}^{\infty }\frac{B_{2n}}{%
2n(2n-1)z^{2n-1}}+\sum\limits_{k=1}^{\infty }\widetilde{S}{_{k}\left( \theta
\right) e^{\pm 2\pi ikz},}  \label{Gammasymp}
\end{equation}%
in the sector $\left\vert \arg z\right\vert \leq \pi -\delta <\pi $ for any $%
0<\delta \leq \pi $ with\footnote{%
Note that, with regards to the location of Stokes lines, that the
asymptotics of the Gamma function is with variable $z$ whereas of the
G-Barnes function is $z+1$.}%
\begin{equation}
\widetilde{S}_{k}\left( \theta \right) =%
\begin{cases}
0 & \text{ if }\left\vert \theta \right\vert <\frac{\pi }{2} \\ 
\frac{1}{2k} & \text{ if }\theta =\pm \frac{\pi }{2} \\ 
\frac{1}{k} & \text{ if }\frac{\pi }{2}<\left\vert \theta \right\vert <\pi ,%
\end{cases}%
,
\end{equation}%
where the usual definition 
\begin{equation*}
\Gamma ^{\ast }(z)=\frac{\Gamma (z)}{\sqrt{\pi }z^{z-1/2}e^{-z}},
\end{equation*}%
was used. In addition to the different decay, the sign difference between $%
\widetilde{S}_{k}\left( \theta \right) $ and $S_{k}\left( \theta \right) $
is crucial with regards to the cancellation of exponentially small terms in
the asymptotics of the observables. Notice also the difference in how the
variable is written in the argument of the functions in (\ref{Gammasymp}).
Thus, the relevant variable indicating a possible Stokes line crossing, in
this case, is then 
\begin{equation}
\widetilde{z}_{1,\pm }=\pm \mathrm{i}\eta +N_{f}/2-N+1.  \label{zGamma}
\end{equation}

\subsection{Exponentially small contributions in the asymptotics}

Taking into account the specific form of the Stokes multipliers parts in (%
\ref{Gasymp}) and (\ref{Gammasymp}), we discuss the appearance of
exponentially small contributions. In all cases, it is due to the Gamma
asymptotics because the part corresponding to the Stokes multipliers in the
G-Barnes asymptotics will cancel out, as we shall see. Since we want to use
the asymptotic result for $z\rightarrow \infty $ we then need to take the $%
\eta \rightarrow \infty $ limit.

There is a small difference between the cases corresponding to \textit{good}
and \textit{ugly} theories, where we will be in the Stokes line if $\eta
\rightarrow \infty $, in comparison to the case of \textit{bad} theories,
where we will be in the Stokes line, or cross it, also for large but finite $%
\eta $. Since we are considering the $\eta \rightarrow \infty $ limit there
are therefore exponentially small contributions in the asymptotics of all
cases.

For example, in the \textit{good} theory case which is self-dual, $N_{f}=2N$
then $z_{1,\pm }=\pm \mathrm{i}\eta $ and we are on the two Stokes lines
always for the two G-Barnes functions in the denominator of (\ref{Zresult}).
We focus only on the exponentially small contributions to the free energy.
That is, in the piece with the Stokes multipliers in (\ref{Gasymp}) for the
two G-Barnes functions $\log G(i\eta +1)+\log G(-i\eta +1)$. We have that
the two set of contributions cancel each other%
\begin{equation*}
F_{N,\mathrm{good}}^{\mathrm{Stokes}}\sim -\frac{1}{2}\sum_{k=1}^{\infty }%
\frac{e^{-2\pi k\eta }-e^{-2\pi k\eta }}{2\pi ik^{2}}=0.
\end{equation*}%
Notice that this result and the one below also holds for finite $N$, however
for the asymptotics of the rest of G Barnes functions in (\ref{Zresult}) one
needs to take the large $N$ limit. Regarding the Gamma asymptotics, using (%
\ref{zGamma}), we have that $\widetilde{z}_{1,\pm }=\pm \mathrm{i}\eta +1$.
Therefore, while for large but finite $\eta $ we have not reached the Stokes
line in this case, the $\eta \rightarrow \infty $ limit effectively puts us
on the Stokes line and we have exponentially small contributions in the
asymptotics, which are those of the Gamma function (\ref{Gammasymp}). We
write down one case explicitly below.

In the \textit{ugly} case $N_{f}=2N-1$ then $z_{1,\pm }=\pm \mathrm{i}\eta
-1/2$ and we have crossed the Stokes line in the G-Barnes function
asymptotics. As above, we have cancellation, due to the symmetric way in
which the $G$-Barnes function appear in the partition function. Focussing on
the piece with the Stokes multipliers in (\ref{Gasymp}), for the two
G-Barnes functions $\log G(i\eta +1/2)+\log G(-i\eta +1/2)$ we have%
\begin{equation*}
F_{N,\mathrm{ugly}}^{\mathrm{Stokes}}\sim -\sum_{k=1}^{\infty }\frac{%
e^{-2\pi k\eta +\pi ik}-e^{-2\pi k\eta -\pi ik}}{2\pi ik^{2}}=0.
\end{equation*}%
In addition, this is the last case where the Gamma function asymptotics
would not contribute at large but finite $\eta $, since the Stokes line is
not yet reached, as $\widetilde{z}_{1,\pm }=\pm \mathrm{i}\eta +1/2$. Again,
regardless of this, the $\eta \rightarrow \infty $ limit situates us in the
Stokes line, leading to exponentially small contributions to the asymptotics.

For bad theories, we have $N_{f}=2N-2-l$ with $l=0,1,2,...$ then $z_{1,\pm
}=\pm \mathrm{i}\eta -1-l/2$ for the G-Barnes function and $\widetilde{z}%
_{1,\pm }=\pm \mathrm{i}\eta -l/2$ for the corresponding Gamma function
asymptotics, then%
\begin{equation*}
F_{N,\mathrm{bad}}^{\mathrm{Stokes}}=-\sum_{k=1}^{\infty }\frac{e^{-2\pi
k\eta }\sin (\pi kl)}{2\pi ik^{2}}-\sum_{k=1}^{\infty }\frac{2e^{-2\pi k\eta
}\cos \left[ 2\pi k\left( 1+l/2\right) \right] }{k}=-2\left( -1\right)
^{l}\sum_{k=1}^{\infty }\frac{e^{-2\pi k\eta }}{k},
\end{equation*}%
and therefore the sign of the contribution is opposite according to the
parity of $N_{f}$. Therefore, we have seen that only the exponentially small
terms in the asymptotics of the Gamma function eventually contribute.
Another way of obtaining this result, would have been to directly invoke the
equivalent expression (\ref{ZGamma}) for the partition function, but it is
illustrative to obtain it from the asymptotics of the G-Barnes function.

Now, localization on $S^{3}$ \cite{Closset:2012vg} permits also to consider
the case $\eta \in 
%TCIMACRO{\U{2102} }%
%BeginExpansion
\mathbb{C}
%EndExpansion
$. If we write $\eta =\eta _{R}+i\eta _{I}$ then, there are clearly many
more possibilities of crossing Stokes lines and the four G-Barnes functions
can now give corresponding exponentially small contributions, because the
crossings are now determined by:%
\begin{eqnarray*}
\pm \eta _{I}+N_{f}/2-N &\leq &0 \\
\pm \eta _{I}+\text{ }N_{f}/2 &\leq &0,
\end{eqnarray*}%
for the Stokes multipliers in (\ref{Gasymp}) and 
\begin{eqnarray*}
\pm \eta _{I}+N_{f}/2-N &\leq &0 \\
\pm \eta _{I}+\text{ }N_{f}/2 &\leq &0,
\end{eqnarray*}%
for the ones in (\ref{Gammasymp}), corresponding to the Gamma function.

Besides, we saw above that this case is equal to an unnormalized (not
divided by the partition function) Wilson loop with a rectangular
representation of the type $(N^{\eta _{I}})$. The case of Wilson loops is
somewhat richer since we have also the additional parameter $r$,
characterizing the representation, but is analyzed in the same way.

\subsection{Duality}

Some simple tests of Seiberg duality can be quickly carried out. This is an
additional test of the analytical formula (\ref{Zresult}). We consider the
generic case given by $N_{f}=2N+k$ for a general integer value of $k$. We
start first with $k=-1$. Then, theory $U(N)$ with $N_{f}=2N-1$ is in the 
\textit{ugly} class, containing a decoupled free sector, generated by BPS\
monopole operators of dimension $\frac{1}{2}$. It is known that the rest is
dual to the IR-limit of the $U(N-1)$ gauge theory with $N_{f}=2N-1$, and
thus a \textit{good} theory. Using (\ref{Zresult}) and again $G(z+1)=\Gamma
(z)G(z)$ one quickly finds that%
\begin{equation}
Z_{N}(N_{f}=2N-1)=e^{-im\eta }N\Gamma (-i\eta +1/2)\Gamma (i\eta
+1/2)Z_{N-1}(N_{f}=2N-1),  \label{uglygood}
\end{equation}%
the $N$ arises due to the fact that we dropped the $N!$ normalizing factor
in (\ref{mod1}) and denoted the resulting partition function by $Z_{N}$,
restoring it, we obtain that for (\ref{mod1}) it holds that%
\begin{equation}
Z_{N_{f}=2N-1}^{U(N)}=Z_{N_{f}=2N-1}^{U(N-1)}\frac{\pi e^{-im\eta }}{\cosh
(\pi \eta )},  \label{d}
\end{equation}%
and we have the expected duality between the $U(N)$ and $U(N-1)$ theories,
together with the expected appearance of a free hypermultiplet. This is the
case where the duality is between a \textit{good} and \textit{ugly} theory,
whereas the rest will be between \textit{good} and \textit{bad} theory.

If we take $k=1$ instead of $k=-1$ then we have to obtain the same duality,
but starting from the good theory side. An immediate computation shows that
this is indeed the case, giving again (\ref{d}). The self-dual case $%
N_{f}=2N $ is evident and the rest is, naively, between $\mathit{good}$ and 
\textit{bad} theories, which corresponds to starting with a $U(N)$ theory
with $N_{f}>2N+1.$ For example, for $U(N)$ theory with $N_{f}=2N+2$, we have%
\begin{equation}
Z_{N}(N_{f}=2N+2)=\frac{e^{2im\eta }}{(N+2)(N+1)}\left\vert \Gamma (-i\eta
+1)\right\vert ^{-2}\left\vert \Gamma (-i\eta )\right\vert
^{-2}Z_{N+2}(N_{f}=2N+2),  \label{goodbad}
\end{equation}%
then%
\begin{equation*}
Z_{N_{f}=2N+2}^{U(N+2)}=Z_{N_{f}=2N+2}^{U(N)}\frac{\pi ^{2}e^{-2im\eta }}{%
\sinh ^{2}(\pi \eta )}.
\end{equation*}%
Thus, this is the duality between a good theory on the l.h.s. of (\ref%
{goodbad}) and a bad theory on the r.h.s. This duality check can be extended
to the generic cases of even and odd number of flavours $N_{f}=2N+2k$ for $%
k=1,2,3,...$ and $N_{f}=2N+2k+1$ for $k=0,1,2,3,...$ (the negative $k$ gives
the same duality as its positive counterpart). We obtain:%
\begin{eqnarray*}
Z_{N_{f}=2N+2k}^{U(N+2k)} &=&Z_{N_{f}=2N+2k}^{U(N)}e^{-2ikm\eta
}\dprod\limits_{j=1-k}^{k}\left\vert \Gamma \left( j+i\eta \right)
\right\vert ^{2}=Z_{N_{f}=2N+2k}^{U(N)}\left( \frac{\pi e^{-im\eta }}{\sinh
(\pi \eta )}\right) ^{2k}, \\
Z_{N_{f}=2N+2k+1}^{U(N+2k+1)} &=&Z_{N_{f}=2N+2k+1}^{U(N)}e^{-i(2k+1)m\eta
}\dprod\limits_{j=-k}^{k}\left\vert \Gamma \left( j+\frac{1}{2}+i\eta
\right) \right\vert ^{2}=Z_{N_{f}=2N+2k+1}^{U(N)}\left( \frac{\pi e^{-im\eta
}}{\cosh (\pi \eta )}\right) ^{2k+1}.
\end{eqnarray*}

\section{Outlook}

We expect to further study the asymptotics together with the duality,
including further discussion on the case of Wilson loops and the setting
where a gauge-R Chern-Simons term is present, characterized by an additional
imaginary part in the FI parameter. The asymptotics in this case will admit
more possibilities and it should be possible to also look at it from the
point of view of Borel transforms. The Mellin-Barnes type of integral given
for the $SU(N)$ theory can also be exploited for both a full analytical
solution and for a study of asymptotics as well.

We conclude by briefly commenting on the fact that the matrix model studied
here not only is related to an extended Selberg integral \cite{FK,FK2}, but
also it has several equivalent representations \cite{FK,FK2}, a well-known
result in the context of Berezin quantization, whose original connections
with matrix models precisely relied heavily on matrix integration identities 
\cite{Hua}. In this way, the matrix model for the partition function without
a FI term also appears in the (Berezin) quantization analysis characterizing
the Hilbert space of a collective field theory of the singlet sector of the
symplectic $Sp(2N)$ sigma model, of importance in the study of dS/CFT
correspondence \cite{Das:2012dt,Anninos:2015eji}. In that setting, it
computes the size of the Hilbert space. In principle the Wilson loop studied
here will also have an interpretation in this Hilbert space picture.

\section*{Acknowledgements}

The author is indebted to Masazumi Honda for many discussions and very
valuable comments and questions. Thanks also to Jorge Russo and David Garc%
\'{\i}a for comments on a preliminary version and to a referee for a useful
observation. This work is supported by the Fund\~{a}\c{c}ao para a Ci\^{e}%
ncia e Tecnologia (program Investigador FCT IF2014), under Contract No.
IF/01767/2014. %\end{acknowledgement}

\end{document}